\documentclass[conference]{IEEEtran}

\usepackage{amsfonts}
\usepackage{graphicx}
\usepackage{amsmath}
\usepackage[numbers, square, comma, sort&compress]{natbib}
\usepackage{amssymb}
\usepackage{hyphenat}
\usepackage{subfigure}
\usepackage{verbatim}
\usepackage{algorithmic}
\usepackage{algorithm}
\usepackage{mathrsfs}
\usepackage{setspace}
\usepackage{booktabs}
\usepackage{longtable}
\usepackage{rotating}
\usepackage{array}
\usepackage{multicol}
\usepackage{multirow}
\usepackage{url}
\usepackage[squaren]{SIunits}
\usepackage{float}
\usepackage{color}
 
\title{Unsupervised Learning in Neuromemristive Systems} 
\author{\IEEEauthorblockN{Cory Merkel and Dhireesha Kudithipudi}
\IEEEauthorblockA{Department of Computer Engineering\\
Rochester Institute of Technology\\
Rochester, New York 14623-5603\\
Email: \{cem1103,dxkeec\}@rit.edu}
}
\begin{document}
\maketitle{}

\begin{abstract}

Neuromemristive systems (NMSs) currently represent the most promising platform to achieve energy efficient neuro-inspired computation.  However, since the research field is less than a decade old, there are still countless algorithms and design paradigms to be explored within these systems.  One particular domain that remains to be fully investigated within NMSs is unsupervised learning.  In this work, we explore the design of an NMS for unsupervised clustering, which is a critical element of several machine learning algorithms.  Using a simple memristor crossbar architecture and learning rule, we are able to achieve performance which is on par with MATLAB's k-means clustering.

\end{abstract}

\section{Introduction}

The present research explores multiple aspects of \textit{unsupervised learning} in a class of neurologically-inspired computer architectures referred to as \textit{neuromemristive systems} (NMSs).  Although they are closely related, NMSs differ from neuromorphic systems--pioneered by Mead in the late 1980s \cite{Mead1990}--in two respects:  First, they are designed using a mixture of CMOS and memristor technologies, which affords levels of connectivity and plasticity that are not achievable in neuromorphic systems.  The second distinction, which is more subtle but equally important, is that NMSs focus on abstraction, rather than mimicking, of neurological processes.  This abstraction generally improves the efficiency of the resulting hardware implementations.

NMS research and development took off rapidly in the late 2000s, coinciding with a growing interest in two-terminal memristors, which will be described briefly in Section \ref{section_memristors} for  unfamiliar readers.  The utility of these systems has been demonstrated in various application domains, especially image processing/analysis.  See \cite{Kudithipudi2014} for a review. Learning in these systems has primarily been supervised.  Although there are some examples of unsupervised learning in spike-based systems \cite{Yu2013}, it is relatively unexplored in non-spiking NMSs.  One example where unsupervised learning in non-spiking networks has been demonstrated is in \cite{Choi2015}, where principal component analysis (PCA) is used to reduce the dimensionality of images.  However, the authors to not discuss any circuit for implementing the PCA algorithm in hardware.

In this research, we propose a non-spiking NMS design for unsupervised clustering.  The NMS is tested on samples from the MNIST database of handwritten digits.  To the best of our knowledge, this is the first work that explores general aspects of clustering in these systems.  Given its ability to reduce data dimensionality and aid in classification, we believe that clustering is a key primitive for future NMSs.  Furthermore, we believe this work will advance the state of unsupervised learning in NMSs and help others do the same.

\section{Brief Description of Memristors}
\label{section_memristors}

Technically, a memristor can be defined as any two-terminal device with a state-dependent Ohm's law \cite{LChua2011}.  More concretely, a memristor is a thin film (I) sandwiched between a top (M) and bottom (M) electrode.  The stack is referred to as a metal-insulator-metal (MIM) structure because the film material is nominally insulating.  That is, in its stoichiometric crystalline form it will have a large band gap and not enough free carriers to conduct.  The film is made conductive by introducing defects in the crystalline structure, either through fabrication, applying an electric field, or both.  Defects may be interstitial metallic ions which are oxidized at one electrode and then drift to the other, where they are reduced.  Defects may also be vacancies such as oxygen vacancies in a TiO$_\text{2}$ film.  In addition, defects may be changes in polarization, such as those in ferroelectric films, or even just changes in crystallinity as in phase change memory.  In some films, the defect profile can be gradually adjusted by applying electric fields for short durations, yielding incremental modifications to the film's overall conductance.  In other films, only two conductance states can be reached.  Moreover, there is usually a minimum amount of energy required to effect change in the film's defect profile.  This often translates to a threshold voltage which must be applied across the film to change its conductance.  Given the constant evolution of memristor technology, it makes little sense to design an NMS around any specific memristor device parameters.  Instead, we assume devices will have these general characteristics:  (1) a large minimum resistance value (e.g. in k$\Omega$s), (2) a large OFF/ON resistance ratio (at least 10\textsuperscript{3}), (3) high endurance (ability to switch many times before failing), (4) high retention (non-volatility), and (5) incremental conductance states that can be reached by applying bipolar voltage pulses above a particular thershold voltage.  All of these properties have been demonstrated in various devices.  See \cite{Kuzum2013} for a review.

\section{Clustering Algorithm Design}
\label{section_clustering}

Clustering algorithms uncover structure in a set of $m$ unlabeled input vectors $\{\mathbf{u}^{(p)}\}$ by identifying $M$ groups, or clusters of vectors that are similar in some way.  In one common approach, each cluster is represented by its centroid, so the clustering algorithm is reduced to finding each of the $M$ centroids.  This can be achieved through a simple competitive learning algorithm:  Initialize $M$ vectors $\mathbf{w}_{i}$ by assigning them to randomly-chosen input vectors.  These will be referred to as weight vectors.  Then, for each input vector, move the closest weight vector a little closer.  After several iterations, the algorithm should converge with the weight vectors lying at (or close to) the centroids.  Of course, there are several parameters which must be defined, including a distance metric for measuring closeness.  The most obvious choice is the $\ell^{2}$-norm.  However, computing this is expensive in terms of hardware because it requires units for calculating squares and square roots.  In addition, as we will discuss later, it is easy to use a high-density memristor circuit called a crossbar to compute dot products between input and weight vectors.  Therefore, it is preferred to use a dot product as a distance metric.  For example, if all of the vectors are normalized ($\lVert\mathbf{u}^{(p)}\rVert=\lVert\mathbf{w}_{i}\rVert=1$), then $\mathbf{w}_{i*}\cdot\mathbf{u}^{(p)}>\mathbf{w}_{i}\cdot\mathbf{u}^{(p)}\forall \mathbf{w}_{i}\ne\mathbf{w}_{i*}$, where $\mathbf{w}_{i*}$ is the closest weight vector to $\mathbf{u}^{(p)}$.  However, the constraint that $\lVert\mathbf{u}^{(p)}\rVert=\lVert\mathbf{w}_{i}\rVert=1$ creates a large overhead, because every input vector has to be normalized and every weight vector has to be re-normalized each time it is updated.

We propose the following solution:  Map each input vector to the vertex of a hypercube centered about the origin:  $\mathbf{u}^{(p)}\in\{-1,1\}^{N}$, where $N$ is the dimensionality of the input space.  Now, $\mathbf{w}_{i}\cdot\mathbf{u}^{(p)}$ will yield a scalar value $d_{i,p}^{*}$ between $-N$ and $+N$.  Moreover, this scalar value can be linearly transformed to a distance $d_{i,p}$ which is the $\ell^{1}$-norm, or Manhattan distance, between the weight vector and the input:
\begin{equation}
d_{i,p}\equiv N-d_{i,p}^{*}=\sum\limits_{j=1}^{N}\lvert w_{i,j}-u_{j}^{(p)}\rvert.
\end{equation}

Using this distance metric, we don't ever need to re-normalize the weight vectors.  Furthermore, mapping input vectors to hypercube vertices can usually be accomplished by thresholding.  For example, grayscale images can be mapped by assigning -1 to pixel values from 0 to 127 and +1 to pixel values from 128 to 255.  Algorithm \ref{algorithm_clustering} summarizes the algorithm.  The first two lines are initialization steps.  Within the double \texttt{for} loop $x_{i}$ is 1 when $i$ corresponds to the index of the closest vector (called the winner) and 0 otherwise.  Then, the weight components of the winner are moved closer to the current input vector using a Hebbian update rule.  The pre-factor $\alpha$, which is called the learning rate, determines how far the weight vectors move each time they win.  Notice that this algorithm is completely unsupervised, so there are no labeled input vectors.

\begin{algorithm}[!t]
\caption{Proposed clustering algorithm.}
\label{algorithm_clustering}
\begin{algorithmic}[1]
\STATE Map inputs to hypercube vertices.
\STATE Initialize weight vectors to random input vectors.
\FOR{$epoch$ = 1:$N_{epochs}$}
  \FOR{$p$ = 1:$m$}
    \STATE $d_{i,p}^{*}=\mathbf{w}_{i}\cdot\mathbf{u}^{(p)}$\hspace{2mm}$\forall i=1,2,\ldots,M$
    \STATE $x_{i}=\begin{cases} 1, & d_{i,p}^{*}=\mathrm{max}(d_{i,p}^{*})\\ 0, &\text{otherwise}\end{cases}\forall i=1,2,\ldots,M$
    \STATE $\Delta w_{i,j} = \alpha x_{i}u_{j}^{(p)}$\hspace{2mm}$\forall i=1,2,\ldots,M$\hspace{2mm}$\forall j=1,2,\ldots,m$
  \ENDFOR
\ENDFOR
\end{algorithmic}
\end{algorithm}

\section{NMS Hardware Design}

\begin{figure}[!t]
\centering
\includegraphics[width=0.75\columnwidth]{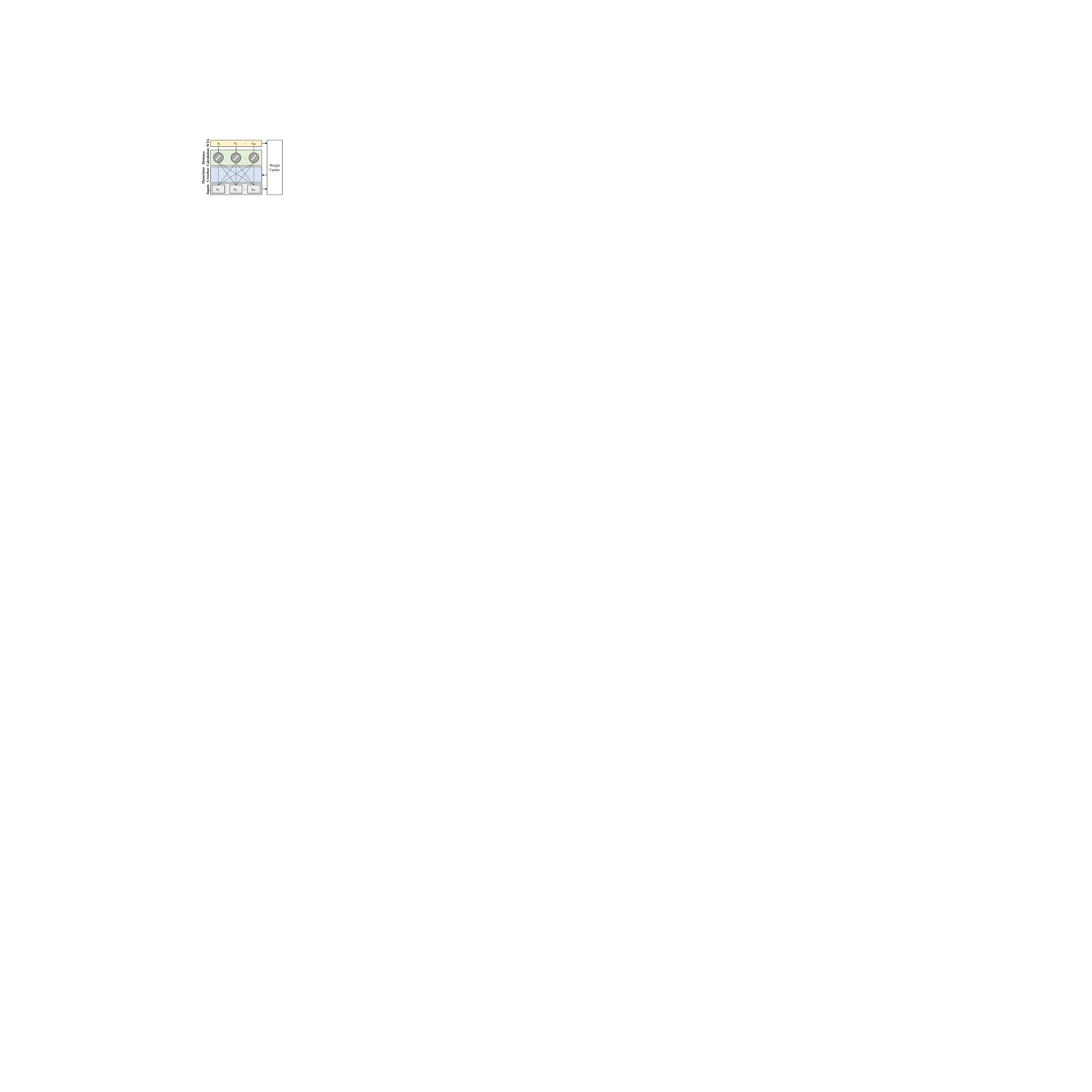}
\caption{Block diagram of proposed NMS for unsupervised clustering.}
\label{figure_clustering_architecture}
\end{figure}

The unsupervised clustering algorithm discussed in Section \ref{section_clustering} can be implemented efficiently in an NMS by representing weight vectors as memristor conductances.  A block diagram of the proposed design is shown in Figure \ref{figure_clustering_architecture}.  The inputs, which are represented as positive and negative currents, are fed through $M$ crossbar circuits.  Together with a non-inverting summing amplifier, (represented as a circle), each crossbar computes the distance between the current input and the weight vector represented by its memristors' conductances.  

\begin{figure}[!t]
\centering
\includegraphics[width=0.95\columnwidth]{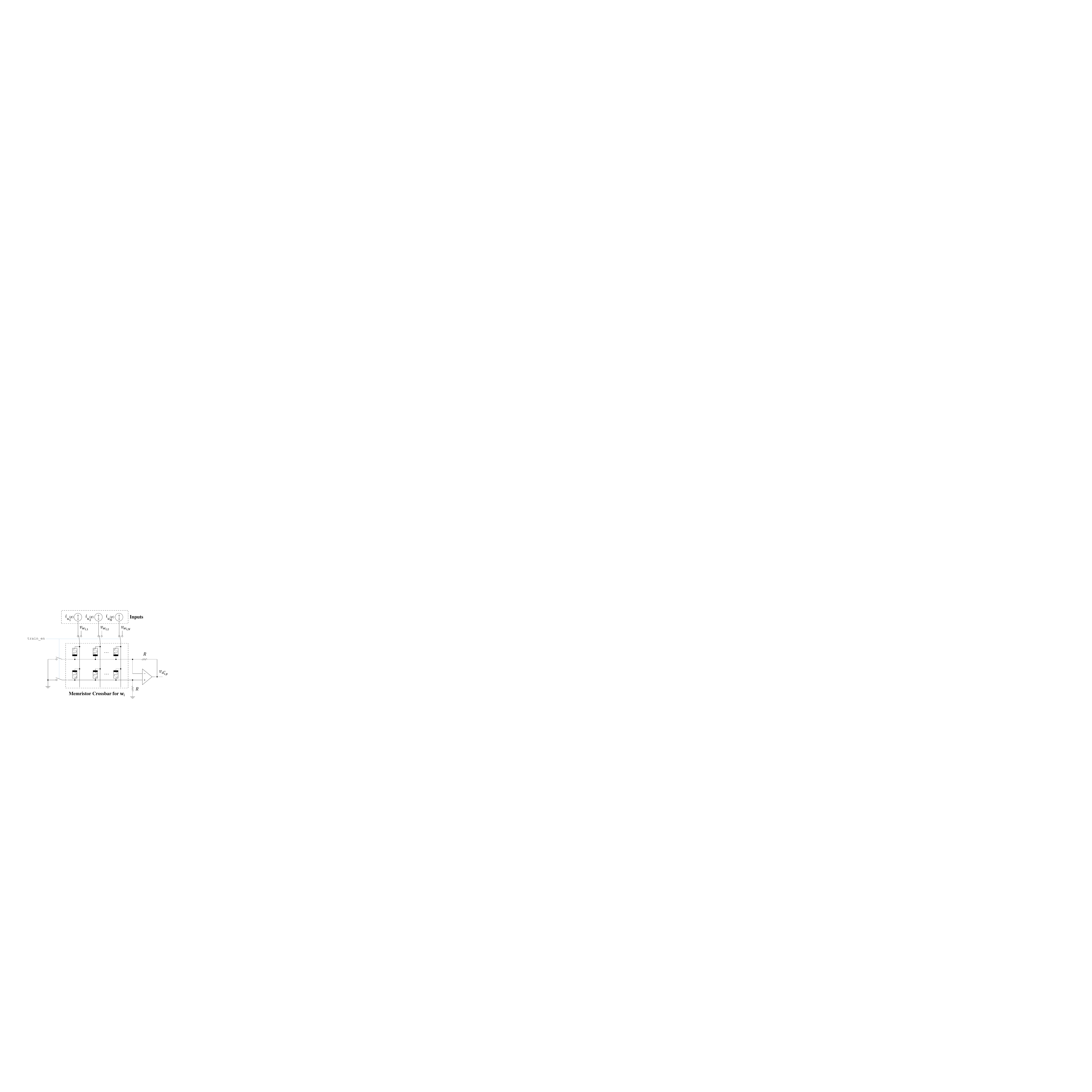}
\caption{Crossbar and summing amplifier circuit for computing the distance between the input and a weight vector.}
\label{figure_crossbar_clustering}
\end{figure}

\begin{figure*}[!t!]
\centering
\includegraphics[width=0.9\textwidth]{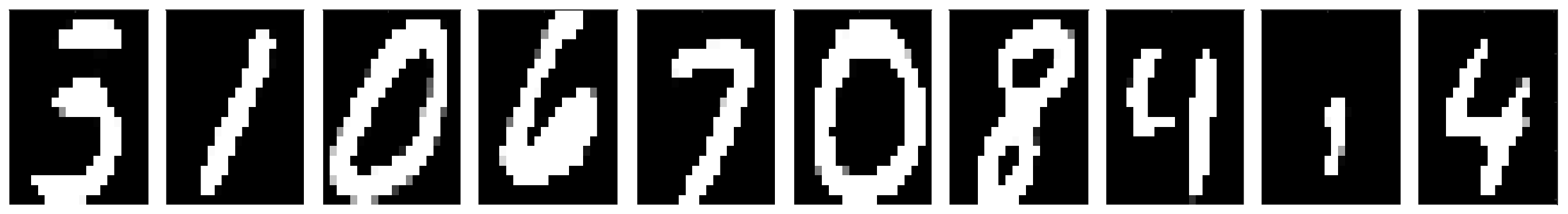}
\caption{10 cluster centroids found in a set of 1000 MNIST images using the proposed NMS.}
\label{figure_hypercube_clustering_mnist_clusters}
\end{figure*}

The configuration of the crossbar and summing amplifier is shown in Figure \ref{figure_crossbar_clustering}.  Memristors in the top row inhibit, or contribute a negative component to the output, while memristors in the bottom row excite, or contribute a positive component to the output.  Therefore, each crossbar column represents one component of one weight vector $\mathbf{w}_{i}$, which can be positive or negative.  If we assume that the op amp has a high open loop gain and the wire resistances are small, then 
\begin{equation}
v_{d_{i,p}^{*}}=\sum\limits_{j=1}^{N}i_{u_{j}^{(p)}}R\left(\frac{G_{2}-G_{1}}{G_{1}+G_{2}}\right)_{i,j},
\end{equation}
where $G_{1}$ and $G_{2}$ are the top and bottom memristors in each column, respectively.  The output of the circuit is a voltage representation of the distance between the current input and the weight vector represented by the crossbar.  The weight vectors are modified by connecting them to write voltages $v_{w_{i,j}}$ using a training enable signal \texttt{train\_en}.  The write voltages are determined by the value of $\Delta w_{i,j}$ in line 7 of Algorithm \ref{algorithm_clustering}.  Specifically, if $\Delta w_{i,j}$ is negative, then $v_{w_{i,j}}$ will be a negative voltage below the memristor's write threshold, and if $\Delta w_{i,j}$ is positive, then $v_{w_{i,j}}$ will be a positive voltage above the memristor's write threshold.  Otherwise, the write voltage is zero. 

So far, we have only discussed the memristor crossbar and distance calculation parts of Figure \ref{figure_clustering_architecture} (line 5 in Algorithm \ref{algorithm_clustering}).  The winner-takes-all circuit (line 6 in Algorithm \ref{algorithm_clustering}) can be implemented in a number of ways.  In this work, we used the current-mode design described in \cite{Lazzaro1988}.  Finally, the weight update (line 7 in Algorithm \ref{algorithm_clustering}) can be computed using simple combinational logic circuits.

\section{Clustering MNIST Images}

\begin{figure}
\centering
\includegraphics[width=0.75\columnwidth]{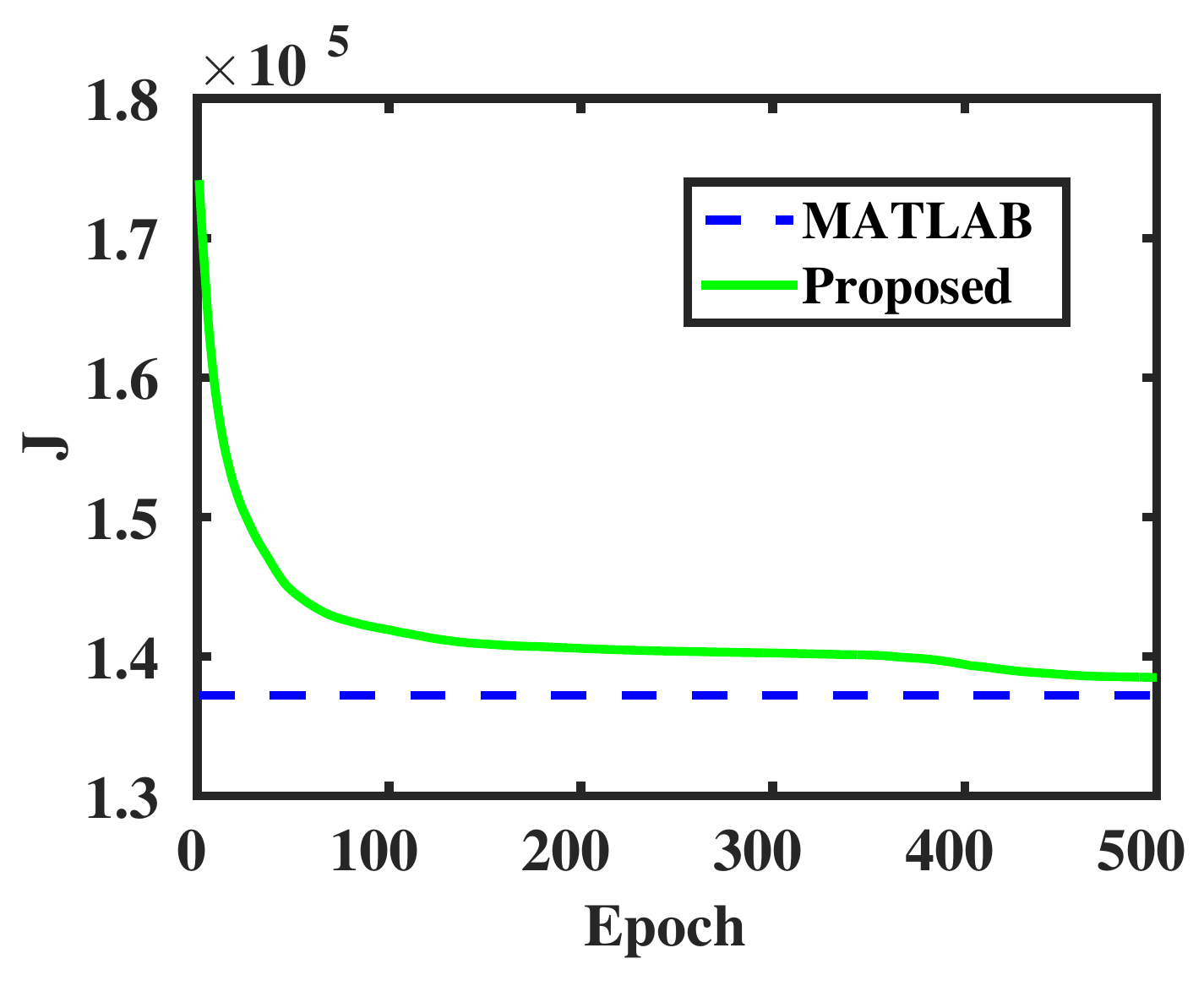}
\caption{Cost function versus epoch while clustering MNIST images using the proposed NMS.}
\label{figure_hypercube_clustering_mnist}
\end{figure}

One exciting application of the proposed hardware is automatically identifying clusters in sets of images.  We took 1000 images ($m$=1000) from the MNIST handwritten digit dataset and clustered them using a behavioral model of the NMS described in the last section.  Each image was originally 20$\times$20 grayscale pixels ($N$=400).  They were mapped to hypercube vertices using the thresholding approach discussed earlier.  In addition, we used 10 clusters ($M$=10), 500 training epochs ($N_{train}$=500), and $\alpha$=0.005.  The results are shown in Figure \ref{figure_hypercube_clustering_mnist_clusters}.  Here, we have plotted the weight vectors representing the centroid of each cluster.  Figure \ref{figure_hypercube_clustering_mnist} shows the cost versus the training epoch, where the cost is defined as
\begin{equation}
J=\sum\limits_{p=1}^{m}\left(\mathrm{min}\hspace{1mm}d_{i,p}\forall i\right).
\end{equation}
We see that the cost function for the proposed NMS approaches that of MATLAB's built-in k-means clustering after 500 epochs.

\section{Conclusions}

The goal of this work was to explore both algorithmic and hardware design aspects of unsupervised learning in NMSs.  To that end, we proposed a clustering algorithm that maps inputs to vertices of a hypercube, and then iteratively finds clusters' centroids using a Hebbian learning rule.  We argue (although we haven't proven) that the proposed algorithm can be implemented more efficiently in an NMS than algorithms that use either $\ell^{2}$-norm or cosine similarity as a distance function.  The algorithm was implemented in a custom NMS design that leverages crossbar circuits to compute the distance between inputs and weight vectors.  To test our design, we clustered 1000 MNIST images and found the results to be consistent with MATLAB's k-means clustering implementation.


\begin{thebibliography}{1}
\providecommand{\url}[1]{#1}
\csname url@samestyle\endcsname
\providecommand{\newblock}{\relax}
\providecommand{\bibinfo}[2]{#2}
\providecommand{\BIBentrySTDinterwordspacing}{\spaceskip=0pt\relax}
\providecommand{\BIBentryALTinterwordstretchfactor}{4}
\providecommand{\BIBentryALTinterwordspacing}{\spaceskip=\fontdimen2\font plus
\BIBentryALTinterwordstretchfactor\fontdimen3\font minus
  \fontdimen4\font\relax}
\providecommand{\BIBforeignlanguage}[2]{{%
\expandafter\ifx\csname l@#1\endcsname\relax
\typeout{** WARNING: IEEEtran.bst: No hyphenation pattern has been}%
\typeout{** loaded for the language `#1'. Using the pattern for}%
\typeout{** the default language instead.}%
\else
\language=\csname l@#1\endcsname
\fi
#2}}
\providecommand{\BIBdecl}{\relax}
\BIBdecl

\bibitem{Mead1990}
C.~Mead, ``{Neuromorphic electronic systems},'' \emph{Proceedings of the IEEE},
  vol.~78, no.~10, pp. 1629--1636, 1990.

\bibitem{Kudithipudi2014}
D.~Kudithipudi, C.~Merkel, M.~Soltiz, G.~S. Rose, and R.~Pino, ``{Design of
  neuromorphic archtectures with memristors},'' in \emph{Network Science and
  Cybersecurity}, R.~Pino, Ed.\hskip 1em plus 0.5em minus 0.4em\relax Springer,
  2014, pp. 93--103.

\bibitem{Yu2013}
S.~Yu, B.~Gao, Z.~Fang, H.~Yu, J.~Kang, and H.-S.~P. Wong, ``{A low energy
  oxide-based electronic synaptic device for neuromorphic visual systems with
  tolerance to device variation.}'' \emph{Advanced Materials}, vol.~25, no.~12,
  pp. 1774--9, Mar. 2013.

\bibitem{Choi2015}
S.~Choi, P.~Sheridan, and W.~D. Lu, ``{Data Clustering using Memristor
  Networks.}'' \emph{Scientific reports}, vol.~5, p. 10492, Jan. 2015.

\bibitem{LChua2011}
L.~Chua, ``{Resistance switching memories are memristors},'' \emph{Applied
  Physics A}, vol. 102, no.~4, pp. 765--783, Jan. 2011.

\bibitem{Kuzum2013}
D.~Kuzum, S.~Yu, and H.-S.~P. Wong, ``{Synaptic electronics: materials, devices
  and applications.}'' \emph{Nanotechnology}, vol.~24, no.~38, p. 382001, Sep.
  2013.

\bibitem{Lazzaro1988}
J.~Lazzaro, S.~Ryckebusch, M.~A. Mahowald, and C.~A. Mead, ``{Winner-take-all
  networks of O(N) complexity},'' in \emph{Advances in Neural Information
  Processing Systems}, 1988, pp. 703--711.

\end{thebibliography}

\end{document}